\documentclass[a4paper,11pt]{article}
\usepackage{pos}
\usepackage{subcaption}

\title{Adressing the p$\Omega^-$ interaction and di-baryonic states via femtoscopy}

\author[a,b]{Marta Piscitelli}
\author*[b]{Otón Vázquez Doce}
\renewcommand{\printHeadAuthors}{Marta Piscitelli}

\affiliation[a]{Department of Physics, Eberhard Karls University of T\"ubingen,\\ Auf der Morgenstelle 10, 72076 T\"ubingen, Germany}

\affiliation[b]{INFN, Laboratori Nazionali di Frascati,\\
Via Enrico Fermi 54, Frascati, Italy}

\emailAdd{marta.piscitelli@student.uni-tuebingen.de}
\emailAdd{Oton.Vazquez.Doce@cern.ch}

\abstract{Motivated by recent experimental measurements of the p$\Omega^-$ correlation function and the concurrent theoretical efforts to describe the strong interaction among hadrons in the strangeness sector, we present a data-driven approach for fine tuning of a meson-exchanges potential for the p$\Omega^-$ system. 
Using femtoscopy data from the ALICE and STAR collaborations, we constrain the strength of the interaction, encoded in the tunable short-range parameter introduced in the potential. The resulting model provides a good description of the measured correlation functions and favors the existence of a bound state in the \({}^5S_2\) channel with a binding energy of approximately \(0.5\,\mathrm{MeV}\). The role of the \({}^3S_1\) channel, however, remains poorly constrained due to the absence of an accurate model accounting for its inelastic contributions. 

}

\FullConference{The 21st International Conference on Hadron Spectroscopy and Structure (HADRON2025)\\
 27 - 31 March, 2025\\
Osaka University, Japan\\}

\begin{document}
\maketitle
\section{Introduction}
The study of hadronic interactions remains one of the key challenges in nuclear physics, particularly in the context of non-perturbative quantum chromodynamics (QCD).
A variety of theoretical approaches have been developed to address this problem, among them phenomenological methods such as effective field theories (EFTs) have proven to be very successful. A major recent step forward came with lattice QCD calculations. In particular, following improvements in both high-performance computing and theoretical methods, the HAL QCD collaboration~\cite{lattice} was the first to provide first-principles interaction potentials from lattice QCD simulations performed at near-physical quark masses (e.g., $m_{\pi} \simeq 146;{\rm{MeV}}/c^2$, $m_{K} \simeq 525;{\rm{MeV}}/c^2$) for several hadron-hadron systems with strangeness content.
In this context the proton-Omega (p$\Omega^-$) system presents an especially interesting case.
The presence of a multistrange baryon places it in the high-strangeness sector ($S = -3$), making it accessible to current lattice QCD calculations due to heavier quark masses. Also very recently, from the experimental point of view, such interactions have been studied via femtoscopy.
This positions the p$\Omega^-$ system at the intersection of robust theoretical predictions and experimental accessibility, making it a particularly promising target for study. 
The system couples to two spin states: the spin-triplet ($^5S_2$), and the spin-singlet ($^3S_1$). Crucially, both lattice QCD and meson-exchange models predict the existence of a shallow bound state in the spin-triplet ($^5S_2$) channel, motivating focused experimental efforts to confirm this theoretical prediction. The spin-singlet ($^3S_1$) channel, in contrast, remains poorly constrained theoretically due to the complexity of modeling its coupling to inelastic channels.

From an experimental point of view, accessing the strong interaction between hadrons in such exotic systems has been made recently possible via femtoscopic techniques through the study of the momentum correlations of particle pairs produced in high-energy collisions. Femtoscopic techniques have been applied to a variety of hadron pairs even reaching the charm sector in the current decade.

The key observable in femtoscopy is the correlation function $ C(k^*)$, for which a theoretical expectation can be built thanks to the Koonin-Pratt formula~\cite{Koonin,Pratt}:
\begin{equation}\label{eq:cf}
    C(k^*) = \int d\mathbf{r}^*\; S(r^*) \; |\psi(\mathbf{k^*}, \mathbf{r^*})|^2 ,
\end{equation}
where $k^* = |\mathbf{k}^*| = |\mathbf{p_2}^* - \mathbf{p_1}^*|/2$ and $r^* = |\mathbf{r}^*|$ denote, respectively, the relative momentum and relative distance of the pair of interest in their rest frame. The quantity $S(r^*)$ represents the so-called source function, describing the distribution of the distance at which the interacting particles are emitted. It is customary to represent the source function with a Gaussian $S(r) = (4\pi r_0)^{-3/2} e^{-(r^2/r_0^2)}$ where $r_0$ defines the source size. The term $|\psi(\mathbf{k^*}, \mathbf{r^*})|$ represents the wave function of the pair, hence containing the interaction part. A recent tool, \emph{Correlation Analysis Tool using the Schrödinger equation} (CATS)~\cite{CATS}, enables the computation of correlation functions through a fully quantum-mechanical solution of the Schrödinger equation for a given interaction potential and source distribution.

The first experimental measurement of the p$\Omega^-$ correlation function was reported by the STAR collaboration~\cite{star} in high-energy heavy-ion collisions, in particular Au-Au collisions measurements at center of mass energy $\sqrt{(s_{NN})} = 200\;\rm{GeV}$. The ALICE collaboration  later showed that the same technique can employed at the LHC~\cite{alice}, where the study of p-p and p-Pb collisions demonstrated to be suited for investigating hadronic interactions.
As a matter of fact, one of the key advantages of p-p and p-Pb collisions at LHC energies is that all hadrons are produced within extremely small space-time volumes, with typical inter-hadron distances of approximately $1\;\rm{fm}$.
Conversely, studies in heavy-ion collision experiments lead to comparatively large hadron pairs sources, with sizes of approximately $3-5\;\rm{fm}$ or even larger, limiting the accessible momentum range, typically below $40\;\rm{MeV/c}$, therefore automatically reducing the sensitivity to all interactions occurring at distances shorter than $1\;\rm{fm}$.

Given the availability of both theoretically motivated interaction potentials for the p$\Omega^-$ system and high-quality experimental  data, this work aims to establish a link between the two.
We adopt a hybrid approach that combines a meson-exchange model with femtoscopic correlation functions measured by the ALICE and STAR collaborations. By tuning the potential to reproduce the measured data, we search for signatures of a bound state, as suggested by previous theoretical studies.
\section{Tuning of a meson exchanges baryon-baryon potential}
The S-wave p$\Omega^-$ system admits two spin configurations: $^5S_2$ and $^3S_1$, where $^{2S+1}L_J$ denotes a state with spin $S$, orbital angular momentum $L$, and total angular momentum $J$. In the following, we focus on the $^5S_2$ channel, as both lattice QCD results and effective Lagrangian approaches are currently limited to this configuration.

Among the phenomenological approaches, a meson exchange model for the $^5S_2$ channel was developed by Sekihara, Kamiya and Hyodo~\cite{sekihara}.
The provided interaction potential combines all the possible long range meson exchange mechanism contributions, denoted $V_A$, $V_B$ and $V_{\rm{box}}$, which are  empirically determined, with the short-range contact term, $V_C \propto c$, where $c$ is an unknown coupling constant. 
The resulting potential is summarized as $V = V_A + V_B + V_C + V_{\rm{box}}$.
However, since the overall potential is nonlocal and thus not well-suited for standard few-body calculations, an equivalent form of the potential in coordinate space is derived in the same work:
\begin{equation}
    V_{p\Omega}^{eq} (r) = \frac{1}{4\pi r} \sum_{n=1}^9 C_n \left(\frac{\Lambda^2}{\Lambda^2 - m_n^2}\right)^2\; \left[e^{-m_n r} - \frac{(\Lambda^2 - m_n^2)r +2\Lambda}{2\Lambda} e^{-\Lambda r}\right],
    \label{eq:local_pot}
\end{equation}
where $\Lambda = 100\;\rm{GeV}$ is a cutoff parameter, $m_n = 100\; n\;\rm{MeV}$ and the parameters $C_n$ are found by taking the real part of the last column in Table V of Ref.~\cite{sekihara}, while the imaginary parts can be neglected, as they are expected to be small in the $J=2$ channel due to p-wave suppression.
The formulation in Eq.~\ref{eq:local_pot} was obtained by requiring the interaction to reproduce the scattering length $7.4\;\pm\; 1.6\;\rm{fm}$ at $t/a=11$, found by the HAL QCD collaboration~\cite{lattice} at nearly physical quark masses. This constraint gives a fitted value of the contact term strength $c = -22.1\;\rm{GeV}^{-1}$. Additionally, the resulting binding energy is found to be of the order of $B = 0.1\;\rm{MeV}$, which, as mentioned above, supports the existence of a shallow bound state in the $^5S_2$ channel. This shows qualitative agreement with lattice QCD calculations, also predicting a shallow bound state with binding energies around $1.54 \;\rm{MeV}$, increasing to $2.46 \;\rm{MeV}$ when Coulomb interaction is included.

As an alternative to fitting the meson exchange model to the lattice QCD potential, we tune the unknown coupling constant $c$ using the femtoscopy data from the ALICE and STAR collaborations. In practical terms, the procedure to tune the potential in Eq.~\ref{eq:local_pot} can be described as follows.

As a first step, we separate the contributions in the equivalent potential coefficients from the contact term from the other terms relative to the long-range interaction. This allow us to isolate the unknown coupling constant and investigate the short-rage dynamics. We then express $c$ as $c = -\beta \;22.1\;\rm{GeV}^{-1}$ where $\beta \in \mathbb{R}$ serves as a free tuning parameter. This gives a modified version of  Eq.~\ref{eq:local_pot} in which the potential separates in two contributions: 
\begin{equation}
    V(\beta) = V_{\rm{long-range}} + \beta\; V_{\rm{C}}.
    \label{eq:tuned_pot}
\end{equation} 

To enable direct comparison with experimental data, we fit the correlation functions, obtained via Eq.~\ref{eq:cf} and constructed with the tuned potential in Eq.~\ref{eq:tuned_pot}, to those measured by the ALICE and STAR collaborations. This requires incorporating both the interaction dynamics and experimental conditions.
The pair wave function, appearing in Eq.~\ref{eq:cf}, is fully determined by the interaction and is therefore consistently implemented for both the ALICE and STAR analyses. In contrast, the source function must account for differences in the collision environments and therefore varies between the two experiments.

In p-p collisions at the LHC, the ALICE collaboration~\cite{alice} reports a well-constrained source size of $r = 0.89 \pm 0.05 \;\rm{fm}$ obtained using data-driven methods based on the resonance source model. In contrast, the STAR collaboration at RHIC does not extract a precise source size due to the more complex geometry of Au–Au collisions. Instead, the interaction is accessed via the ``small to large'' (SL) ratio, taking the ratio of correlation functions from peripheral ($40$–$80\%$) to central ($0$–$40\%$) Au–Au collisions, $\frac{C_{0\text{–}40\%}}{C_{40\text{–}80\%}}$, assuming source sizes of $2-3\;\rm{fm}$ and $3-5\;\rm{fm}$, respectively. This approach reduces both Coulomb effects and source-size uncertainties, providing a practical probe of the strong interaction under complex collisions conditions.
Following STAR's approach we adopt the SL-ratio method for the STAR data and compute the ratio of the two correlation functions assuming source sizes $r = 2.5\pm0.5\;\rm{fm}$ for peripheral and $r = 4\pm1\;\rm{fm}$ for central collisions.

In addition, to enable direct tuning of the correlation functions to the experimental data, we introduce two specific normalization parameters to account for differences in background subtraction methods.
Overall the fitted correlation functions can be schematically written as:
\begin{align}
    C^{\rm{ALICE}}_{\rm{fitted}} &= C_{R_{\rm{p-}\Omega} = 0.89\,\rm{fm}}(n_{\rm{ALICE}}, \beta, k^*) = n_{\rm{ALICE}} \, C_{R_{\rm{p-}\Omega} = 0.89\,\rm{fm}}(\beta, k^*) \\
    C^{\rm{STAR}}_{\rm{fitted}} &= n_{\rm{STAR}} \frac{C_{r_0 = 2.5\,\rm{fm}}}{C_{r_0 = 4\,\rm{fm}}}(\beta, k^*)
\end{align}
where $n_{\mathrm{ALICE}}$ and $n_{\mathrm{STAR}}$ are the introduced free normalization parameter for the ALICE and STAR data respectively.

Thus far, only the ${}^5S_2$ channel has been considered. To account for a possible contribution from the ${}^3S_1$ channel, we consider two limiting scenarios: a purely elastic and an inelastic case, following the recipe in Ref.~\cite{akira}. In the purely elastic case, the radial part of the wave function $\chi(r)$ is assumed to be identical for both spin states, i.e., $\chi_{J=1}(r) = \chi_{J=2}(r)$. In the inelastic case, the strong interaction is modeled as an absorptive core represented by a complex potential, $V(r) = -i V_0 \theta(r_0 - r)$, with $V_0 \rightarrow +\infty$ and $r_0 = 2\;\mathrm{fm}$, leading to complete suppression of the wave function: $\chi_{J=1}(r) = 0$. Under these two assumptions, the spin-averaged correlation function can be expressed as: $C_{p\Omega}(k^*) = \frac{5}{8}C_{J=2}(k^*) + \frac{3}{8}C_{J=1}(k^*)$ where $C_{J=1}(k^*)$ differs in the two presented scenarios.
\begin{figure}[htbp]
  \centering


    \includegraphics[width=0.89 \textwidth]{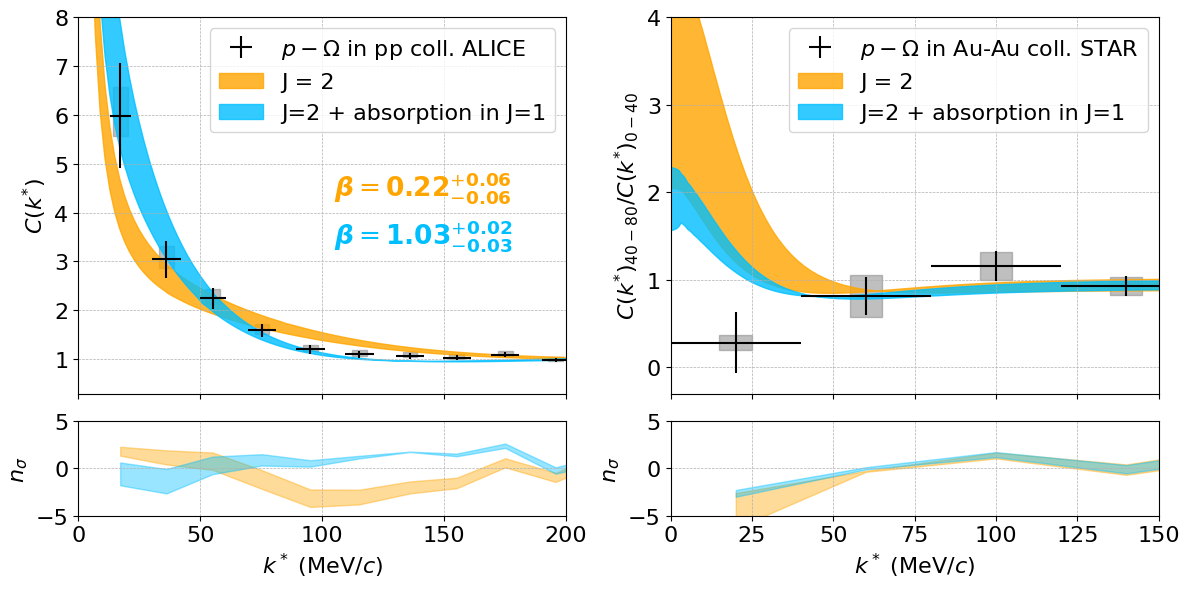}

  \caption{Correlation functions \( C(k^*) \) for the p$\Omega^-$ system measured by ALICE (left) and STAR (right), compared with the fitted correlation functions derived from the tuned potential. Experimental data are shown in black. The bands represent the fitted models under the elastic (orange) and inelastic (blue) scenarios and include a \( 1\sigma \) uncertainty width. See text for details.}
  \label{fig:cf_fit}
\end{figure}

In Fig.~\ref{fig:cf_fit}, we present the correlation functions $C(k^*)$ for the p$\Omega^-$ system, measured by the ALICE and STAR collaborations, together with the fitted correlation functions $C^{\rm{ALICE}}_{\rm{fitted}}$ and C$^{\rm{STAR}}_{\rm{fitted}}$. In the latter, the strength of the short-range interaction, tuned via the free parameter $\beta$, is fitted simultaneously to both datasets. The left and right panels correspond to the ALICE and STAR data, respectively. The black points represent the experimental data, with vertical error bars denoting statistical uncertainties, grey boxes indicating systematic uncertainties, and horizontal bars showing the statistical uncertainty on the mean $k^*$ in each bin. The fitted correlation functions, $C^{\rm{ALICE}}_{\rm{fitted}}$ and C$^{\rm{STAR}}_{\rm{fitted}}$, are shown as colored bands: the orange band corresponds to the assumption of elastic contribution in the $J=2$ channel, while the blue band includes inelastic absorption effects in the $J=1$ channel.
The width of the bands corresponds to the $1\sigma$ interval, which includes uncertainties from the fitted $\beta$ value, the source size, and the systematic uncertainty associated with the data estimated via a bootstrap procedure.

The results of the simultaneous fit under the two limiting scenarios for the $J=1$ channel yield a value for the fitted parameter of $\beta = 0.22^{+0.06}_{-0.06}$ for the purely elastic case and $\beta = 1.03^{+0.02}_{-0.03}$ for the case including full inelastic absorption.
On the other hand, the normalization parameters are fitted independently for each experimental data set, as they are experiment-specific. The fitted values are $n_{\mathrm{ALICE}} = 1.14^{+0.02}_{-0.02}$  and $n_{\mathrm{STAR}} =  0.96^{+0.05}_{-0.05}$ in the inelastic scenario and $n_{\mathrm{ALICE}} = 0.97^{+0.02}_{-0.02}$ and $n_{\mathrm{STAR}} = 0.95^{+0.07}_{-0.07}$ in the purely elastic scenario.

It can be seen that the fitted model provides an accurate description of the correlation functions measured by the ALICE and STAR collaborations, demonstrating the ability of the constructed model to capture the underlying interaction dynamics across different collision systems. 
However, the description of the data is, as expected, highly dependent on the assumptions made for the behavior of the $J=1$ channel. In fact, a good fit to the ALICE data, particularly in the momentum region around 100 MeV/c, where a characteristic depletion associated with the presence of a bound state could be expected, can only be achieved in the inelastic scenario, i.e., by assuming total absorption in the $J=1$ channel. 
It is also worth noting the tension with the lowest momentum bin in the STAR data. It is natural that the fit reproduces the ALICE data more accurately, given its significantly better precision. However, the discrepancy may also stem from the uncertainty in the source size for the Au–Au collisions, and from the fact that STAR does not provide the mean relative momentum within each bin. This mean is typically offset from the bin center (especially for the low-momentum bins, where it tends to shift to higher values) which could lead to a better agreement between the model and the data.

Finally, by solving the Schr\"odinger equation for the potential used in this analysis, we derive the scattering length and binding energy. Neglecting Coulomb effects, a bound state emerges when the parameter $\beta$ exceeds the critical value $\beta = 0.92$, at which the inverse scattering length vanishes ($1/a_0 = 0$), marking the threshold for the formation of a bound state.
As noted before, the fitted values of $\beta$ are found to be highly sensitive to the treatment of the coupled-channels in the $J=1$ channel, marking that the existence and properties of the bound state are strongly influenced by the modeling of these effects. In the limit where the $J=1$ is assumed to be fully absorptive, the fitted value $\beta = 1.03^{+0.02}_{-0.03}$ exceeds the threshold for the formation of a bound state in the $J=2$ channel, resulting in an estimated binding energy of $B \sim 0.5\;\rm{MeV}$. In contrast, for the purely elastic case, the corresponding $\beta$ value remains below the threshold, and no bound state is supported. This result supports the existence of a bound state in the ($^5S_2$) p$\Omega^-$ system, consistent with earlier theoretical predictions based on both lattice QCD~\cite{lattice} and meson-exchange models~\cite{sekihara}.

\section{Conclusions and Outlook}
We have presented a data-driven analysis of the p$\Omega^-$ system using femtoscopy data from the ALICE and STAR collaborations to constrain a meson-exchange potential model. By tuning the short-range component of the interaction, our model achieves a good description of the experimentally measured correlation functions, highlighting the utility of femtoscopy data in refining theoretical models of hadron interactions.

Our results support the existence of a bound state in the $J=2$ channel, in agreement with previous theory predictions. However, definitive conclusions remain limited by significant uncertainties in the $J=1$ channel. In particular, the outcome of the analysis is highly sensitive to the assumptions on coupled-channel effects. Additional femtoscopy measurements involving, for example, direct measurement of the $\Lambda \Xi^-$ correlations, for which an exploratory measurement exists already~\cite{LambdaXi}, could help constrain these effects and clarify the role of the $J=1$ channel interaction. Moreover, measurements of the p$\Omega^-$ correlation function in systems with varying source sizes could shed light on the characteristic depletion pattern associated with the presence of a bound state. In this context, collisions involving systems larger than pp but with less complex dynamics than heavy-ion systems like Pb–Pb or Au–Au would be especially valuable. The current light-ion runs at the LHC with O–O and Ne–Ne collisions, as well as the STAR isobar program, could thus provide particularly useful data.

Further developments of the here presented approach may involve extending our model either by introducing additional tunable parameters to the potential or applying machine learning techniques to more efficiently explore the parameter space, particularly necessary for multi-variate analyses. In parallel, higher-precision femtoscopy measurements across different collision systems, with varying source sizes, will be crucial in resolving the question of the p$\Omega^-$ bound state's existence.

\section*{Acknowledgements}
We would like to thank Takayasu Sekihara, Yuki Kamiya, and Tetsuo Hyodo for the guidance in the use of the interaction potential, for the calculations of the binding energies and for valuable discussions.
\begingroup 

\providecommand{\href}[2]{#2}\begingroup\raggedright\endgroup

\endgroup
\end{document}